\documentclass[letterpaper,12pt]{article}
\usepackage{bm}
\usepackage{tabularx} 
\usepackage{amsmath}  
\usepackage{graphicx} 
\usepackage[margin=1in,letterpaper]{geometry} 
\usepackage{cite} 
\usepackage[final]{hyperref} 
\hypersetup{
	colorlinks=true,       
	linkcolor=blue,        
	citecolor=blue,        
	filecolor=magenta,     
	urlcolor=blue         
}

\begin{document}

\title{Further study of $f_0(1710)$ with coupled-channel approach and hadron molecular picture}
\author{Zheng-Li Wang$^{1,2,}$\footnote{Email address:
  \texttt{wangzhengli@itp.ac.cn} }~   and
  Bing-Song Zou$^{1,2,3}$\footnote{Email address:
  \texttt{zoubs@itp.ac.cn} }
  \\[2mm]
  {\it\small$^1$CAS Key Laboratory of Theoretical Physics, Institute
  of Theoretical Physics,}\\
  {\it\small  Chinese Academy of Sciences, Beijing 100190,China}\\
  {\it\small$^2$School of Physics, University of Chinese Academy of Sciences (UCAS), Beijing 100049, China} \\
  {\it\small$^3$School of Physics and Electronics, Central South University, Changsha 410083, China} \\
}
\date{\today}
\maketitle

\begin{abstract}
The $f_0(1710)$ was previously proposed to be dynamically generated state by interactions between vector mesons. We extend the study of $f_0(1710)$ by including its coupling to channels of pseudoscalar mesons within coupled-channel
approach. The channels involved are $K^*\bar{K}^*,\rho\rho,\omega\omega,\phi\phi,
\omega\phi,\pi\pi,K\bar{K},\eta\eta$. We show that the pole assigned to $f_0(1710)$
does not change much. Then we calculate
the partial decay widths of $f_0(1710) \to K^*\bar{K}^* \to \pi\pi,K\bar{K},\eta\eta$ as the coupled channel dynamically generated state as well as assuming it to be pure $K^*\bar{K}^*$ molecule. In both
cases the ratios of partial decay widths agree fairly with that in PDG.
\end{abstract}

\section{Introduction}
More and more hadron resonances have been proposed to be hadron molecules~\cite{Guo:2017jvc} with much more predicted ones to be searched for~\cite{Dong:2021juy}. Among various approaches for studying hadron molecules, a quite popular one is the unitary extension of chiral perturbation
theory, which has been successfully to study the meson-baryon and meson-meson interactions
at low energy~\cite{Oller:1999ag,Oller:2000fj,Dobado:1996ps,Oller:1998zr,
Oller:1998hw,Oller:1997ti,Oller:2000ma,Molina:2020hde}. 
A well-known example is the $\Lambda(1405)$~\cite{Jido:2003cb}, 
which can be dynamically
generated in the vicinity of the $\pi\Sigma$ and $K^-p$ thresholds. The another
example is $f_0(980)$~\cite{Oller:1997ti,Janssen:1994wn}, 
which is considered to arise due to $\pi\pi$ and $K\bar{K}$ coupled channel
interaction. Some recent works~\cite{Geng:2008gx,Du:2018gyn}
studied the interaction of the nonet of vector mesons themselves
and found a pole with quantum number $J^{PC}=0^{++}$ mainly coupling to $\bar{K}^*K^*$ channel, possibly corresponding to
$f_0(1710)$.

In this paper, we extend the previous study~\cite{Du:2018gyn} of $f_0(1710)$ by including its coupling to channels of pseudoscalar mesons in addition to vector mesons to see how these more coupled channels influence the result on the $f_0(1710)$ pole and meanwhile whether its corresponding partial decay widths to these channels of pseudoscalar mesons compatible with experimental data.   
Our work is organized as follows. In Sect.~\ref{sec:Formalism}, we outline the formalism to the coupled-channel interaction
\cite{Oller:2019opk}.
In Sect.~\ref{sec:Results}, we give our numerical results and discussion with a brief summary at the end.

\section{Formalism}\label{sec:Formalism}
\begin{figure}[htp]
  \centering
  \includegraphics[scale=0.5]{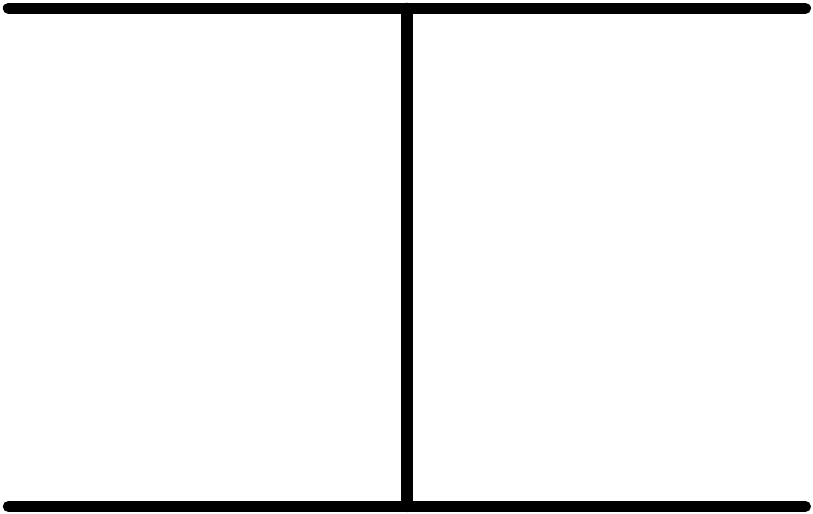}
  \qquad
  \includegraphics[scale=0.5]{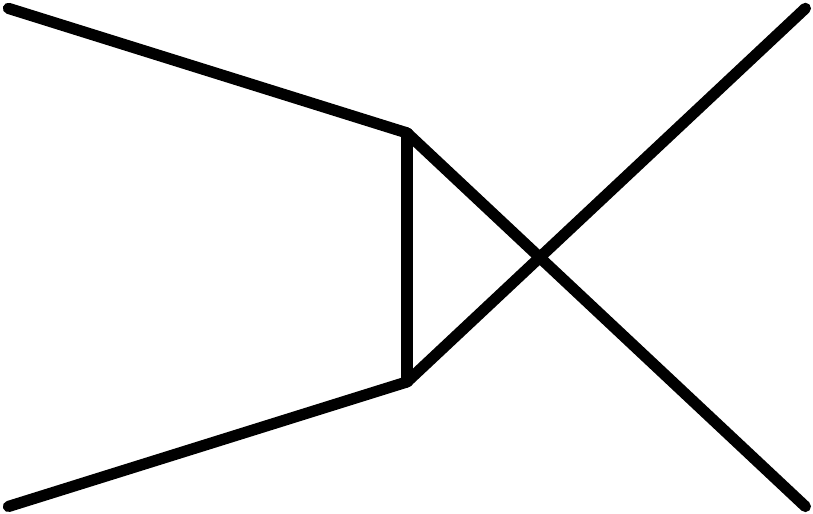}
  \caption{The $t$- and $u$-channel Feynman diagrams} \label{fig:feyndiag}
\end{figure}

The interaction Lagrangian among vector mesons and pseudoscalar mesons is given by
~\cite{Bando:1984ej,Bando:1987br}
\begin{equation} \label{eq:lag}
  \mathcal{L}_{VPP} = -ig \langle V^\mu [P,\partial_\mu P] \rangle.
\end{equation}
where the symbol $\langle \ldots \rangle$ stands for the trace in the $SU(3)$ space
and the coupling constant $g=M_V/2f_\pi$ with $f_\pi = 92MeV$ the pion decay constant.
The vector field $V^\mu$ is
\begin{equation}
  V^\mu = \left(
  \begin{array}{ccc}
    \frac{1}{\sqrt{2}} \rho^0 + \frac{1}{\sqrt{2}} \omega & \rho^+ & K^{*+} \\
    \rho^- & -\frac{1}{\sqrt{2}} \rho^0 + \frac{1}{\sqrt{2}} \omega & K^{*0} \\
    K^{*-} & \bar{K}^{*0} & \phi \\
  \end{array} \right)^\mu,
\end{equation}
and the pseudoscalar field $P$ is
\begin{equation}
  P = \left(
  \begin{array}{ccc}
    \frac{1}{\sqrt{2}} \pi^0 + \frac{1}{\sqrt{6}} \eta & \pi^+ & K^+ \\
    \pi^- & -\frac{1}{\sqrt{2}} \pi^0 + \frac{1}{\sqrt{6}} \eta & K^0 \\
    K^- & \bar{K}^0 & -\frac{2}{\sqrt{6}} \eta \\
  \end{array} \right).
\end{equation}
With the Lagrangian given in Eq.~\eqref{eq:lag}, we are able to calculate the
vector-vector to pseudoscalar-pseudoscalar scattering amplitudes. The Feynman diagrams
needed are shown in Fig.~\ref{fig:feyndiag}. 

The amplitudes with isospin-$0$ for the processes
$V(p_1)V(p_2) \to P(p_3)P(p_4)$ are listed in Table~\ref{tab:potential}. 
\begin{table}[htp]
  \centering
  \[ \begin{array}{cc}
    \hline
    \text{Channel} & T^{(0)} \\
    \hline
    \rho\rho \xrightarrow{\pi} \pi\pi & -16(V^\pi_t + V^\pi_u) \\
    \rho\rho \xrightarrow{K} K\bar{K} & -2\sqrt{6}(V^K_t + V^K_u) \\
    \omega\omega \xrightarrow{K} K\bar{K} & 2\sqrt{2}(V^K_t + V^K_u) \\
    \phi\phi \xrightarrow{K} K\bar{K} & 4\sqrt{2}(V^K_t + V^K_u) \\
    \omega\phi \xrightarrow{K} K\bar{K} & -4(V^K_t + V^K_u) \\
    K^*\bar{K}^* \xrightarrow{K} \pi\pi & -2\sqrt{6}(V^K_t + V^K_u) \\
    K^*\bar{K}^* \xrightarrow{K} \eta\eta & 6\sqrt{2}(V^K_t + V^K_u) \\
    K^*\bar{K}^* \xrightarrow{\pi,\eta} K\bar{K} & -6(V^\pi_t + V^\eta_t) \\
    \hline
  \end{array} \]
  \caption{The potential of each channel with isospin-$0$} \label{tab:potential}
\end{table}
The convention used to relate the particle basis to the isospin basis is
\begin{equation} \begin{split}
  |\pi^+ \rangle &= -|1,1\rangle \qquad |K^+ \rangle = -|\frac12,\frac12 \rangle, \\
  |\rho^+ \rangle &= -|1,1\rangle \qquad |K^{*+} \rangle = -|\frac12,\frac12 \rangle.
\end{split} \end{equation}
The $V_t$ and $V_u$ correspond to the $t$- and $u$-channel diagrams, respectively. The superscript
is the particle exchanged. Here, $t=(p_1-p_3)^2$ and $u=(p_1-p_4)^2$ are the usual
Mandelstam variables. The potential has the form
\begin{equation}
  V_{t(u)}^{ex} =\frac{g^2}{t(u)-m^2_{ex}} \epsilon_1 \cdot p_3 \epsilon_2 \cdot p_4.
\end{equation}
where the $\epsilon_i$ is the $i$-th polarization vector of the incoming vector meson.
The polarization vector can be characterized by its three-momentum $\mathbf{p}_i$ and
the third component of the spin in its rest frame, and the explicit expression of the
polarization vectors can be found in Appendix A of Ref.~\cite{Gulmez:2016scm}.

In term of these amplitudes with isospin-$0$, we can get the $S$-wave potential via
\cite{Gulmez:2016scm}
\begin{equation} \label{eq:projection} \begin{split}
  T^{(JI)}_{\ell S;\bar{\ell} \bar{S}}(s) &=
  \frac{Y^0_{\bar{\ell}}(\hat{\mathbf{z}})}{\sqrt{2}^N (2J+1)}
  \sum_{\begin{subarray}{c}
    \sigma_1,\sigma_2,\bar{\sigma}_1 \\
    \bar{\sigma}_2,m
  \end{subarray}}
  \int \mathrm{d} \hat{\mathbf{p}}'' Y^m_\ell (\mathbf{p}'')^*
  (\sigma_1 \sigma_2 M | s_1 s_2 S) \\
  &\times (mM\bar{M} | \ell SJ) 
  (\bar{\sigma}_1 \bar{\sigma}_2 \bar{M} | \bar{s}_1 \bar{s}_2 \bar{S})
  (0\bar{M} \bar{M} | \bar{\ell} \bar{S} J) \\
  &\times T^{(I)}(p_1,p_2,p_3,p_4;\epsilon_1,\epsilon_2,\epsilon_3,\epsilon_4).
\end{split} \end{equation}
with $s=(p_1+p_2)^2$ the usual Mandelstam variable, $M=\sigma_1+\sigma_2$ and
$\bar{M} = \bar{\sigma}_1 + \bar{\sigma}_2$. And $N$ accounts for the identical
particles, for example
\begin{flalign}
  N &= 2 \text{ for } \rho\rho \to \pi\pi, & \\
  N &= 1 \text{ for } \rho\rho \to K\bar{K}, & \\
  N &= 0 \text{ for } \omega\phi \to K\bar{K}. &
\end{flalign}
Like vector scattering $ VV \to V V$, the partial wave projection
Eq.~\eqref{eq:projection} for a $t$-channel exchange amplitude of
$V V \to P P$ would also develop a left-hand cut via~\cite{Du:2018gyn}
\begin{equation} \begin{split}
  \frac12 & \int_{-1}^{+1} \mathrm{d} \cos \theta \frac{1}{t-m^2_{ex}+i\epsilon} =
  -\frac{s}{\sqrt{\lambda (s,m_1^2,m_2^2) \lambda (s,m_3^2,m_4^2)}} \\
  &\times \log \frac{m_1^2 + m_2^2 - \frac{(s+m_1^2-m_2^2)(s+m_3^2-m_4^2)}{2s} -
  \frac{\sqrt{\lambda (s,m_1^2,m_2^2) \lambda (s,m_3^2,m_4^2)}}{2s} -m_{ex}^2 +i\epsilon}
  {m_1^2 + m_2^2 - \frac{(s+m_1^2-m_2^2)(s+m_3^2-m_4^2)}{2s} +
  \frac{\sqrt{\lambda (s,m_1^2,m_2^2) \lambda (s,m_3^2,m_4^2)}}{2s} -m_{ex}^2 +i\epsilon}.
\end{split} \end{equation}
with $\lambda (a,b,c) = a^2+b^2+c^2-2ab-2bc-2ac$ the K\"all\'en function. In vector
scattering $V V \to V V$, left-hand cuts are smoothed by the $N/D$ method~\cite{Chew:1960iv,Bjorken:1960zz}. 
As for the scattering $V V \to P P$,
all left-hand cuts are located below the $PP$ threshold, which are far away from
the energy region we are interested in, so we do not deal with these cuts.

The basic equation to obtain the unitarized $T$-matrix is
\begin{equation}
  T^{(JI)} (s) = \left[ 1-V^{(JI)}(s) \cdot G(s) \right]^{-1} \cdot V^{(JI)}(s).
\end{equation}
Here $V^{(JI)}$ denotes the partial-wave amplitudes and $G(s)$ is a diagonal matrix
made up by the two-point loop function $g_i(s)$,
\begin{equation}
  g_i(s) \to i \int \frac{\mathrm{d}^4 q}{(2\pi)^4} \frac{1}
  {(q^2-m^2_{i1}+i\epsilon) ((P-q)^2 -m^2_{i2} +i\epsilon)}.
\end{equation}
with $P^2=s$ and $m_{i1,2}$ the masses of the particles in the $i$-th channel. The pole
position is at the zeros of determinant
\begin{equation}
  \text{Det} \equiv \text{det} \left[ 1- V^{(JI)}(s) \cdot G(s) \right].
\end{equation}
The above loop function is logarithmically divergent and can be calculated with a
once-subtracted dispersion relation or using a regularization $f_\Lambda (q)$
\begin{equation}
  g_i(s) = i \int \frac{\mathrm{d}^4 q}{(2\pi)^4} \frac{f^2_\Lambda (q)}
  {(q^2-m^2_{i1}+i\epsilon) ((P-q)^2 -m^2_{i2} +i\epsilon)}.
\end{equation}
after the $q^0$ integration is performed by choosing the contour in the lower half of
the complex plane, we get
\begin{equation}
  g_i(s) = \int_0^\infty \frac{|\mathbf{q}|^2 \mathrm{d}|\mathbf{q}|}{(2\pi)^2}
  \frac{\omega_{i1} + \omega_{i2}}{\omega_{i1}\omega_{i2}
  (s-(\omega_{i1} + \omega_{i2})^2 + \mathrm{i}\epsilon)} f_\Lambda^2(|\mathbf{q}|).
\end{equation}
where $\mathbf{q}$ is the three-momentum and
$\omega_{i1,2} = \sqrt{\mathbf{q}^2 + m^2_{i1,2}}$. In order to proceed we need
to determine $f_\Lambda (\mathbf{q})$. There are two kinds of choices, sharp cutoff
and smooth cutoff, typically:
\begin{equation}
  f_\Lambda (\mathbf{q}) = \left\{
    \begin{array}{l}
      \Theta(\Lambda^2 - \mathbf{q}^2) \\
      \exp\left[-\frac{\mathbf{q}^2}{\Lambda^2}\right]\\
    \end{array} \right.
\end{equation}

In order to compare with the previous results of coupled channel approach~\cite{Du:2018gyn}, the same sharp cutoff is used in this paper when channels with pseudoscalar mesons are included in addition. To explore the
position of the poles we need to take into account
the analytical structure of these amplitudes in the different Riemann sheets. By
denoting $q_{on}$ for the CM tri-momentum of the particles $1$ and $2$ in the 
$i$-th channel
\begin{equation}
  q_i^\text{on} = \frac{\sqrt{(s-(m_{i1}-m_{i2})^2) (s-(m_{i1}+m_{i2})^2)}}{2\sqrt{s}}.
\end{equation}
As the quantity is two-valued itself~\cite{Doring:2009yv}, 
we need to distinguish the two Riemann sheets
of $q_i^\text{on}$ uniquely according to
\begin{equation}
  q_i^{\text{on}>} = \left\{
    \begin{array}{cl}
      -q_i^\text{on} & \text{if } \text{Im} q_i^\text{on} < 0 \\
      q_i^\text{on} & \text{else} \\
    \end{array}\right.
\end{equation}
And the analytic continuation to the second Riemann sheet is  given by 
\begin{equation}
  g^{(2)}_i (s) = g_i (s) + \frac{i}{4\pi} \frac{q_i^{\text{on}>}}{\sqrt{s}}.
\end{equation}

\section{Numerical results and discussion }\label{sec:Results}

First we assume that $f_0(1710)$ is $K^*\bar{K}^*$ hadron molecule state and calculate the partial
decay widths of $f_0(1710) \to K^*\bar{K}^* \to \pi\pi,K\bar{K},\eta\eta$ with the hadronic triangle loop approach~\cite{Du:2019pij,Lin:2019qiv,Guo:2019fdo,Lin:2017mtz}. 

\begin{figure}
  \centering
  \includegraphics[scale=0.5]{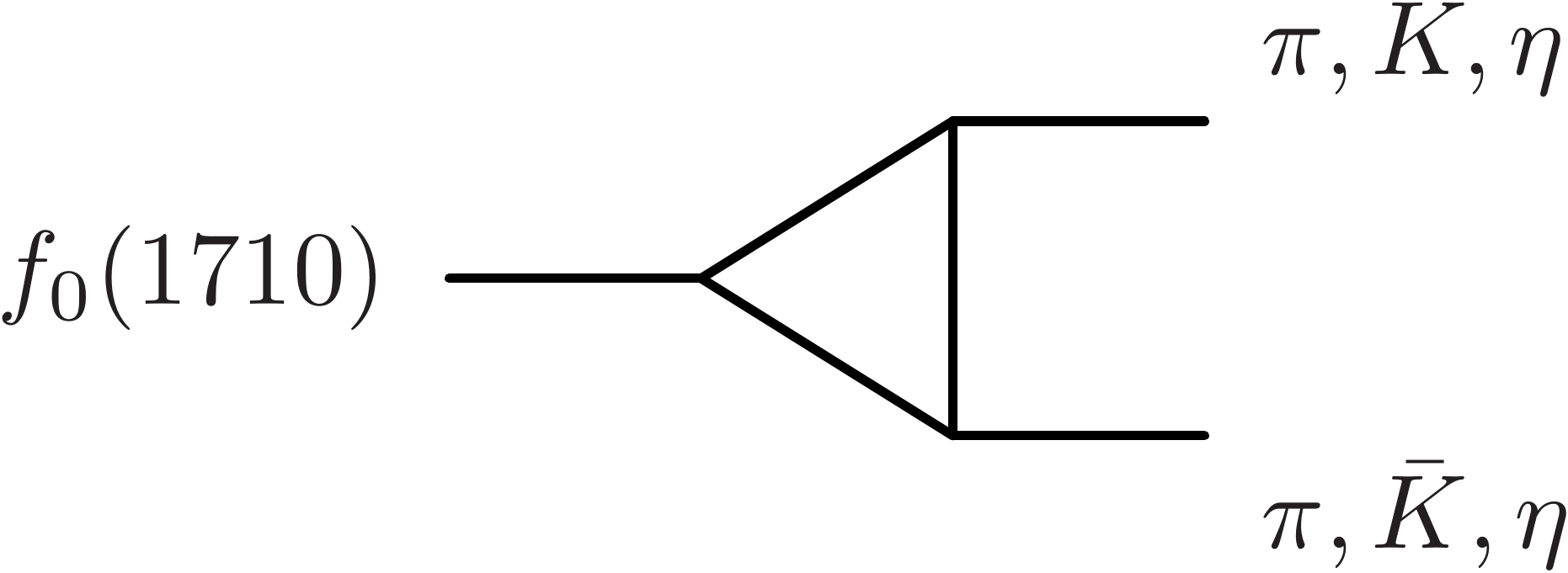}
  \caption{Decay of $f_0(1710)$}\label{fig:decay}
\end{figure}

The loop function corresponding to this process as shown in Fig.~\ref{fig:decay} is 
\begin{equation}
  D = -i\int \frac{\mathrm{d}^4q}{(2\pi)^4}
  \frac{1}{(p_3+q)^2-m^2_{K^*}+i\epsilon}
  \frac{1}{(p_4-q)^2-m^2_{\bar{K}^*}+i\epsilon}
  \frac{1}{q^2-m^2_3+i\epsilon}.
\end{equation}
where for different final state $m_3$ can be the mass of $\pi,K,\eta$. Since the mass of
$f_0(1710)$ is close to the threshold of $K^*\bar{K}^*$, the internal lines with
$K^*$ and $\bar{K}^*$ exchange can be approximated non-relativistically. And the the 
loop function can be simplified as
\begin{equation} \begin{split}
  D &= -i\int \frac{\mathrm{d}^4q}{(2\pi)^4} \frac{1}{4m^2_{K^*}}
  \frac{1}{(p^0-\omega_1)^2-(q^0)^2-i\epsilon} 
  \frac{1}{(q^0)^2-\omega^2_3+i\epsilon} \\
  &\times \left(\frac{\Lambda^2-m^2_{ex}}{\Lambda^2-q^2}\right)^2
  \exp \left[-\frac{(\mathbf{p}+\mathbf{q})^2}{\Lambda^2_G}\right].
\end{split} \end{equation}
with $p_3=(p^0,\mathbf{p}),q=(q^0,\mathbf{q})$ and 
$\omega_1 = \sqrt{(\mathbf{p}+\mathbf{q})^2+m^2_{K^*}}$,
$\omega_3 = \sqrt{\mathbf{q}^2+m^2_3}$. Here for the vertex $f_0(1710) K^* \bar{K}^*$
the Gaussian form factor is added. For the t-channel meson exchange, coupling constants with off-shell meson 
are dressed by monopole form factors~\cite{Machleidt:1989tm,Titov:2000bn}. For simplicity, we take $\Lambda_G$ and 
$\Lambda$ to be equal to be in the range of $0.8\sim 1.0 GeV$. And the results are
\begin{align}
  \frac{\Gamma(f_0(1710)\to \pi\pi)}{\Gamma(f_0(1710\to K\bar{K}))}
  &=0.394\pm 0.134\; (0.23\pm 0.05), \\
  \frac{\Gamma(f_0(1710)\to \eta\eta)}{\Gamma(f_0(1710\to K\bar{K}))}
  &=0.239\pm 0.057\; (0.48\pm 0.15).
\end{align}
where the values of the PDG~\cite{PDG} are given between brackets.
It seems that the calculated partial decay width of $f_0(1710)\to \pi\pi$ is larger than the central value in
PDG, but it is still within the range of large error bar.

\begin{figure}
  \centering
  \includegraphics[scale=0.5]{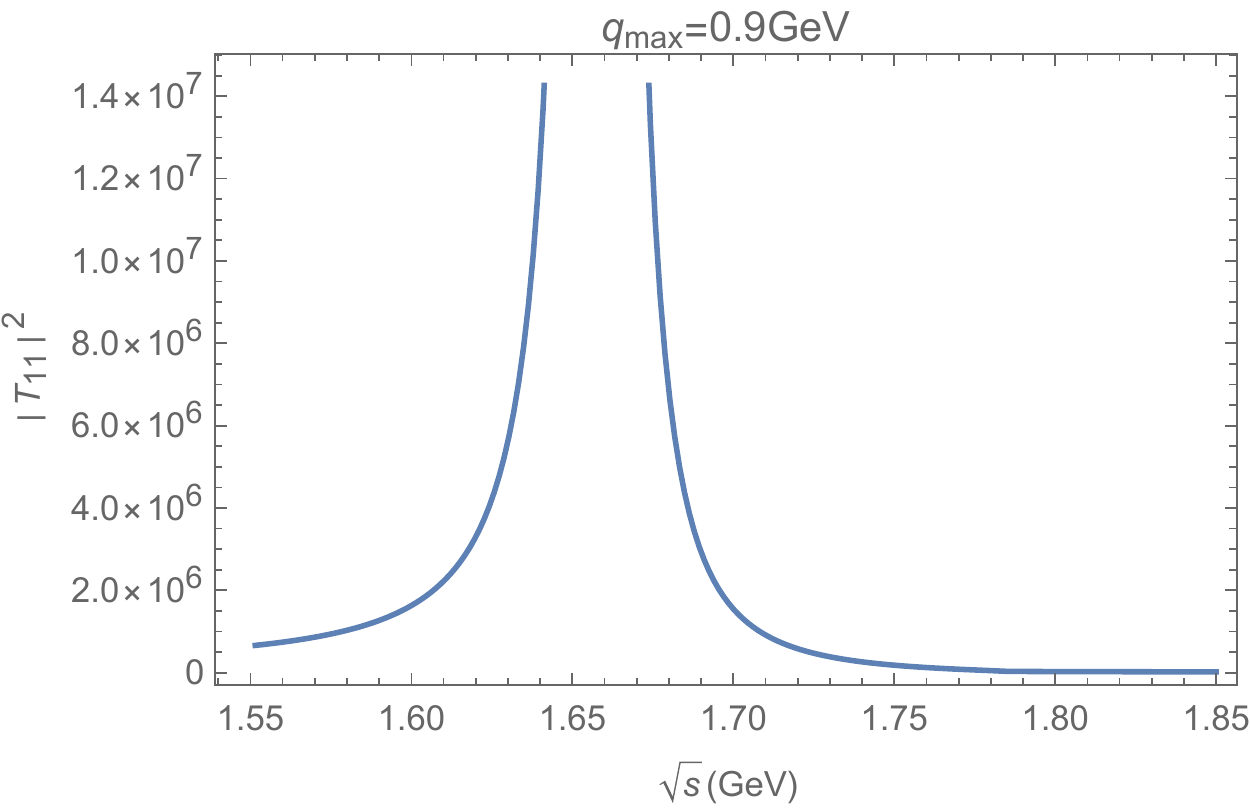}
  \qquad
  \includegraphics[scale=0.5]{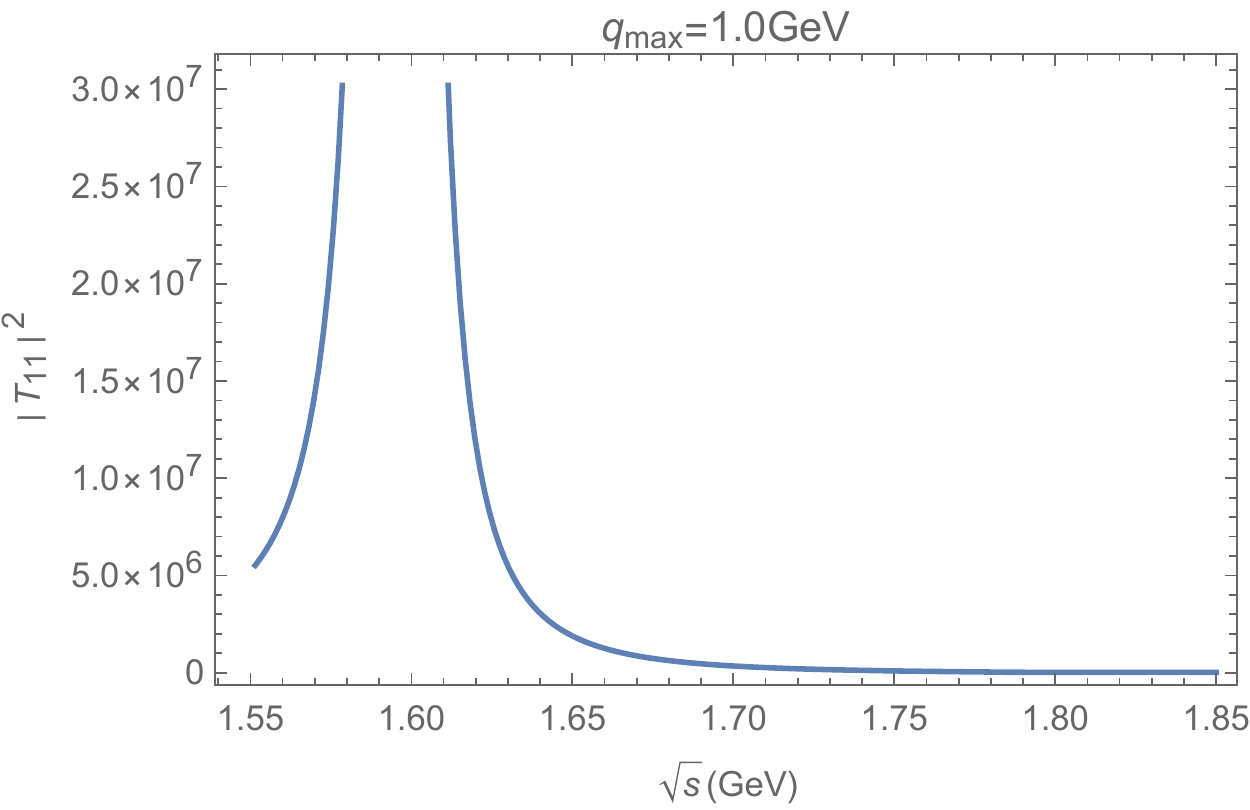}
  \caption{$|T_{11}|^2$ for different $q_{max}=0.9,1.0 GeV$}\label{fig:t11}
\end{figure}

To check how the $f_0(1710)$ is influenced by various coupled channels in the unitary coupled channel approach,
we start with the single $K^* \bar{K}^*$ channel case by dropping its couplings to all other channels.
The Fig.~\ref{fig:t11} gives the $|T|^2$ matrix of scattering 
$K^*\bar{K}^* \to K^*\bar{K}^*$. We label the $K^*\bar{K}^*$ as channel $1$, and 
the remain channel indices are listed in Table~\ref{tab:index}. A bound state is found  
locating at $\sqrt{s}=1.66 GeV$ for cutoff $q_{max}=0.9GeV$ and $\sqrt{s}=1.60 GeV$
for $q_{max}=1.0 GeV$. The bound state moves down when the cutoff $q_{max}$ increases.
\begin{table}
  \centering
  \[\begin{array}{ccc}
    \hline
    \text{Channel index} & \text{Channel} & \text{Threshold}~(GeV) \\
    \hline
    1 & K^*\bar{K}^* & 1.784 \\
    2 & \rho\rho & 1.54 \\
    3 & \omega\omega & 1.564 \\
    4 & \phi\phi & 2.04 \\
    5 & \omega\phi & 1.802 \\
    6 & K\bar{K} & 0.99 \\
    7 & \pi\pi & 0.276 \\
    8 & \eta\eta & 1.096 \\
    \hline
  \end{array}\]
  \caption{Channel indices and threshold energies}\label{tab:index}
\end{table}

Then we turn on additional channels to study their influence on the mass and width of the resonance. The
$T$-matrix for a single channel $a$ is given by 
\begin{equation}
  T_{aa} = \frac{V_{aa}}{1-V_{aa}g_a}.
\end{equation}

If we turn on another channel $b$, then the $T$-matrix for $a\to a$ becomes to be
\begin{equation}
  T_{aa} = \frac{V_{aa}+\frac{V_{ab}^2g_b}{1-V_{bb}g_b}}
  {1-(V_{aa}+\frac{V_{ab}^2g_b}{1-V_{bb}g_b})g_a}.
\end{equation}

Compared to the single channel, $V_{aa}$ is replaced by
\begin{equation}
  V_{eff} = V_{aa}+\frac{V_{ab}^2g_b}{1-V_{bb}g_b}.
\end{equation}

\begin{figure}
  \centering
  \includegraphics[scale=0.4]{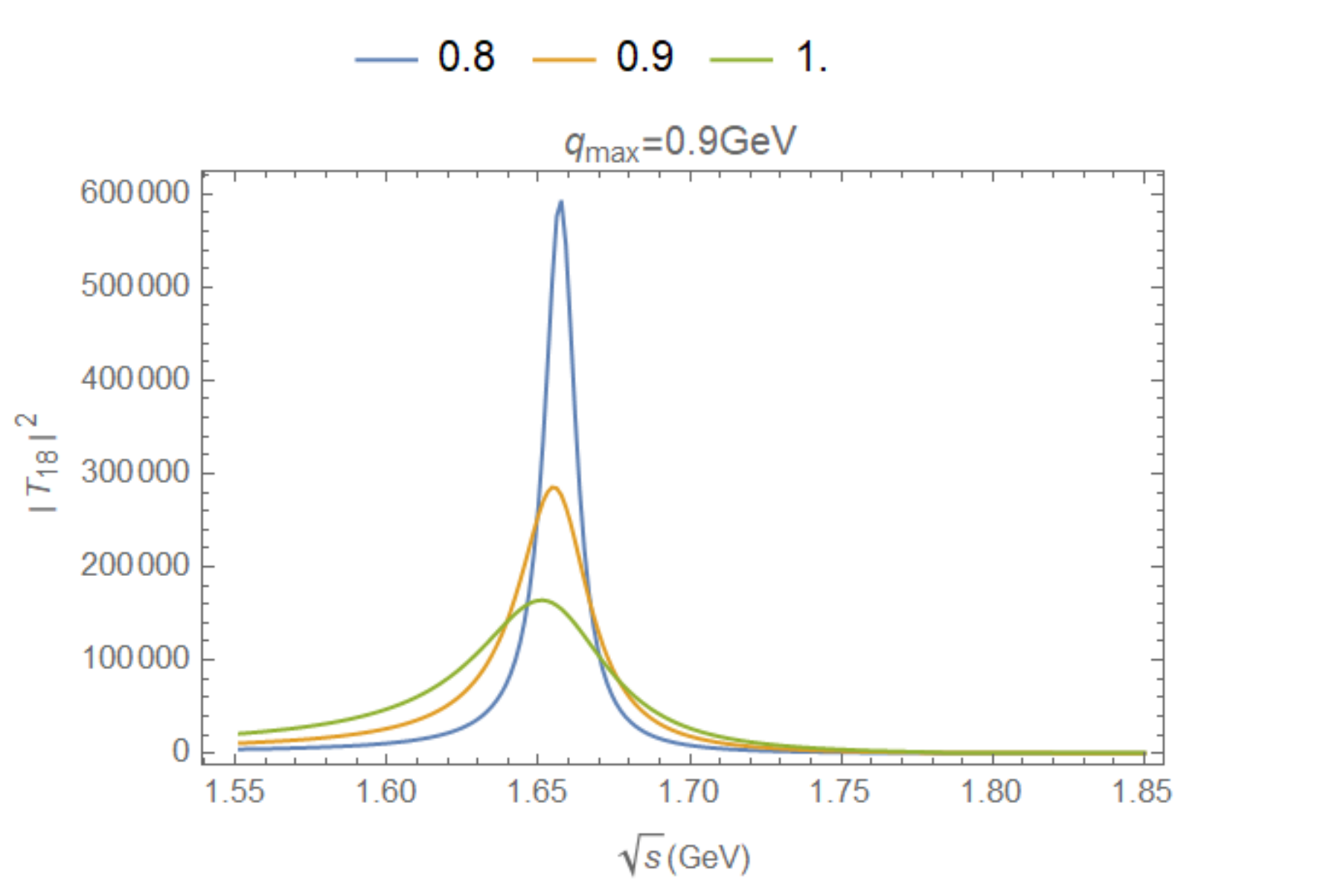}
  \includegraphics[scale=0.4]{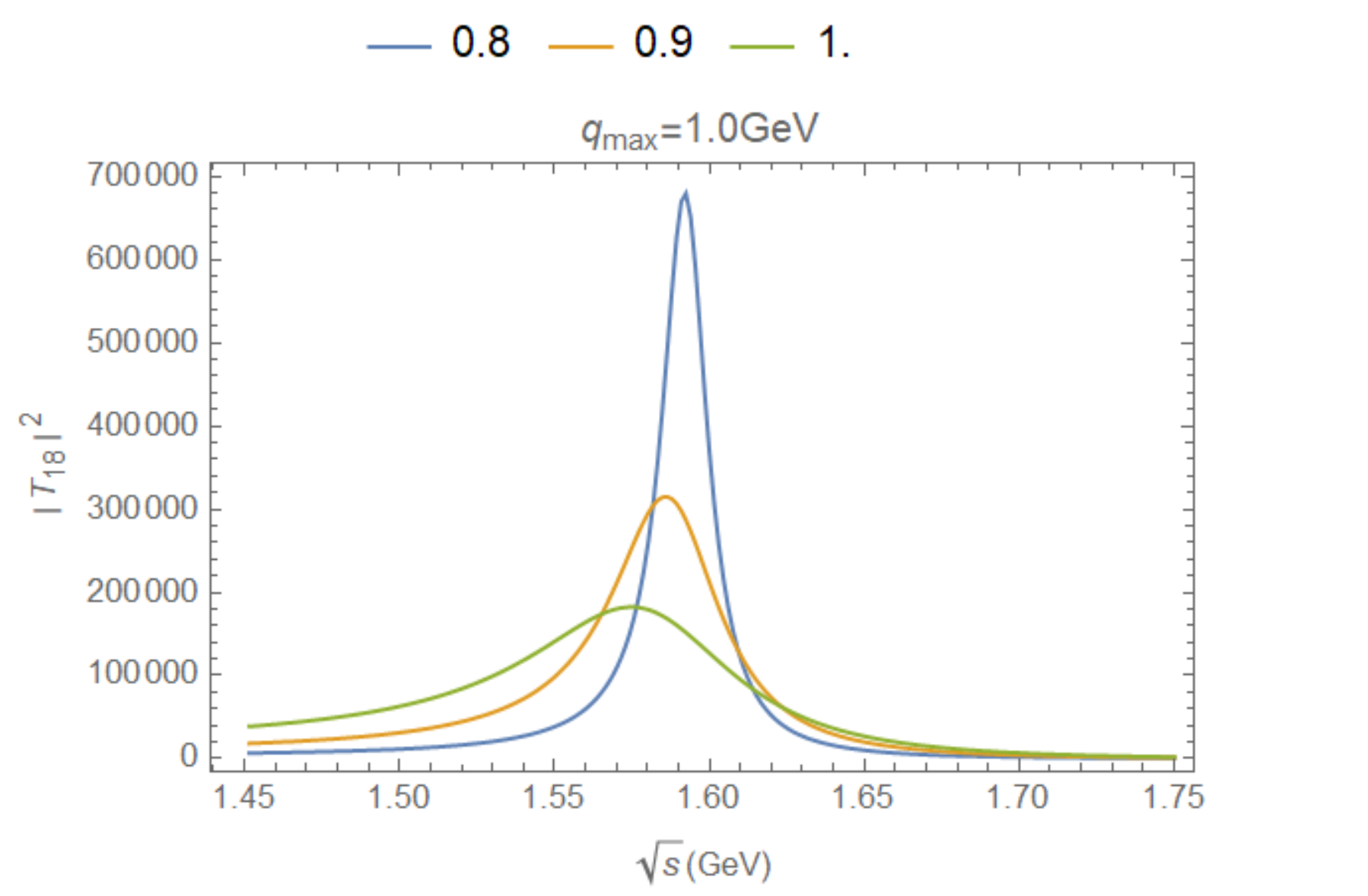}
  \caption{$|T_{18}|^2$ for different $\Lambda=0.8,0.9,1.0 GeV$ and
   $q_{max}=0.9,1.0 GeV$}\label{fig:t18}
\end{figure}

Denoting the second term as
\begin{equation}
  V' = \frac{V_{ab}^2g_b}{1-V_{bb}g_b},
\end{equation}
then $T_{aa}$ can be written as
\begin{equation}
  T_{aa} = \frac{V_{aa}+V'}{1-(V_{aa}+V')g_a}
  = \frac{(1+\alpha)V_{aa}}{1-(1+\alpha)V_{aa}g_a}.
\end{equation}
with $\alpha = V'/V_{aa}$. For calculating the loop integral $g_a$ for channels of vector mesons, we use the same sharp cutoff as in the previous study~\cite{Du:2018gyn}. However, if we turn on the channels of pseudo-scalar mesons, such as $\eta\eta$ channel, the imaginary part of $V'$ is too large compared with the result of triangle loop approach. In order to get consistent results with two approaches, the same kind of monopole form factor of the form 
\begin{equation}
  F = \frac{\Lambda^2 - m^2_{ex}}{\Lambda^2-q^2}.
\end{equation}
is needed at each $VPP$ vertex for the exchanged pseudo-scalar meson with momentum $q$.
When the form factor is implemented, the $T$ matrix for
$K^*\bar{K}^* \to \eta\eta$ is showed in Fig.~\ref{fig:t18}.
For cutoff $q_{max}=0.9 GeV$, the real part of the resonance is around $1.66 GeV$,
which is the same as the single channel. The imaginary part is about
$6,14,26 MeV$ for different $\Lambda=0.8,0.9,1.0 GeV$. And for cutoff $q_{max}=1.0 GeV$,
the situation is similar.

\begin{figure}
  \centering
  \includegraphics[scale=0.5]{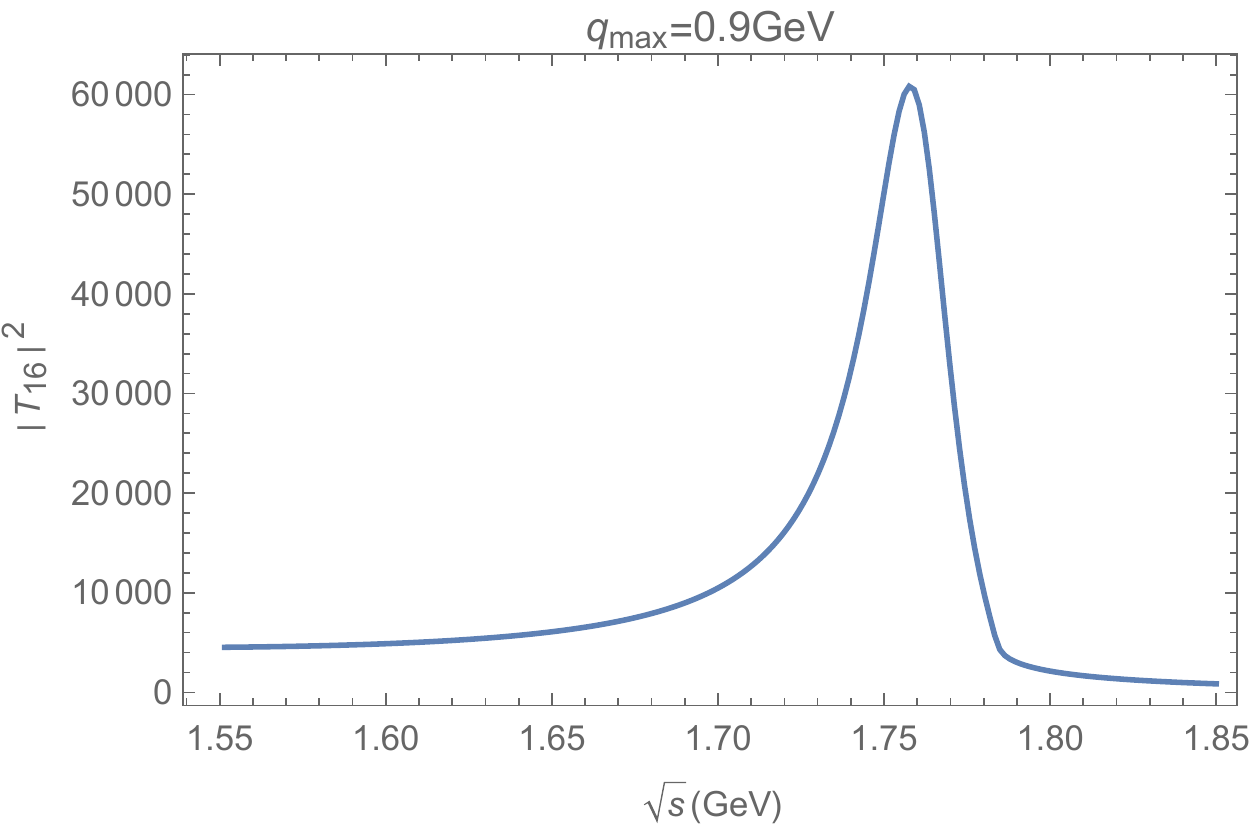}
  \qquad
  \includegraphics[scale=0.5]{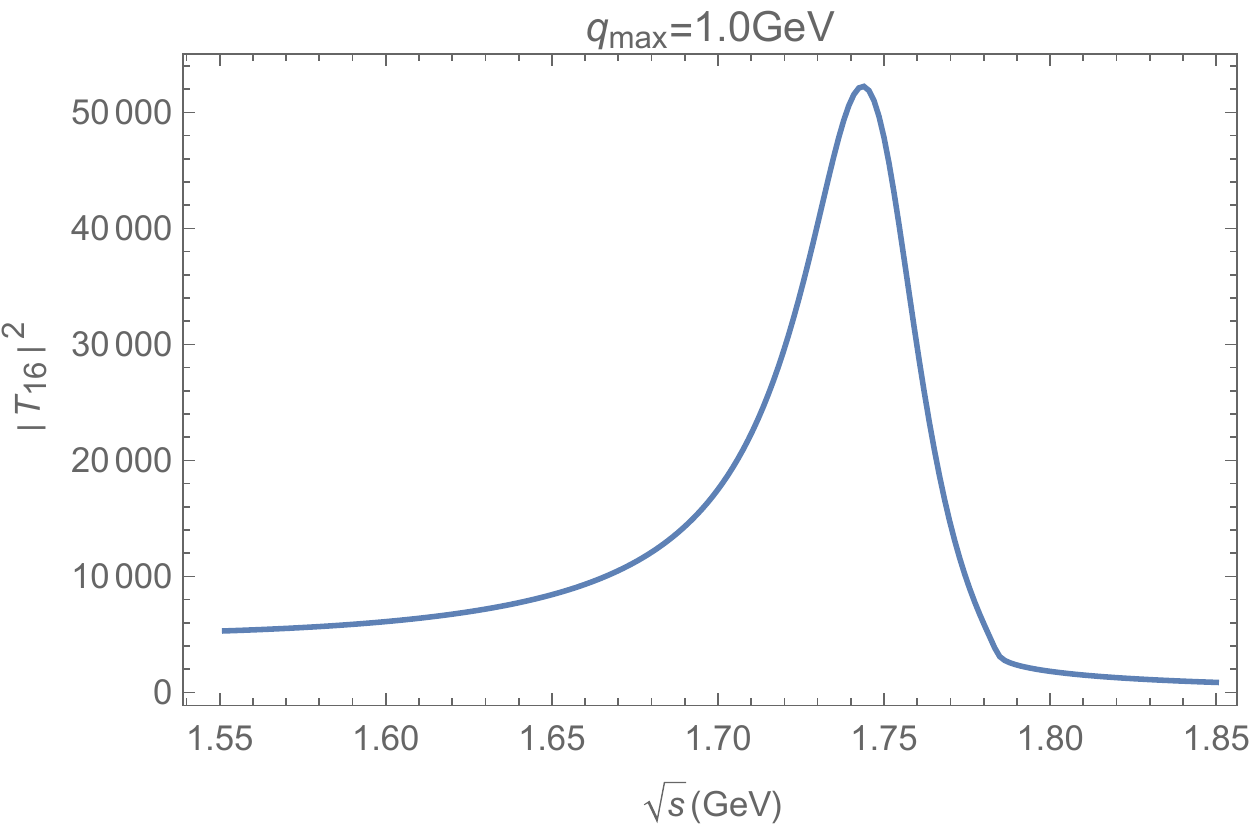}
  \caption{$|T_{16}|^2$ for $\Lambda=0.9 GeV$ and different $q_{max}=0.9,1.0 GeV$}\label{fig:t16}
\end{figure}

\begin{figure}
  \centering
  \includegraphics[scale=0.5]{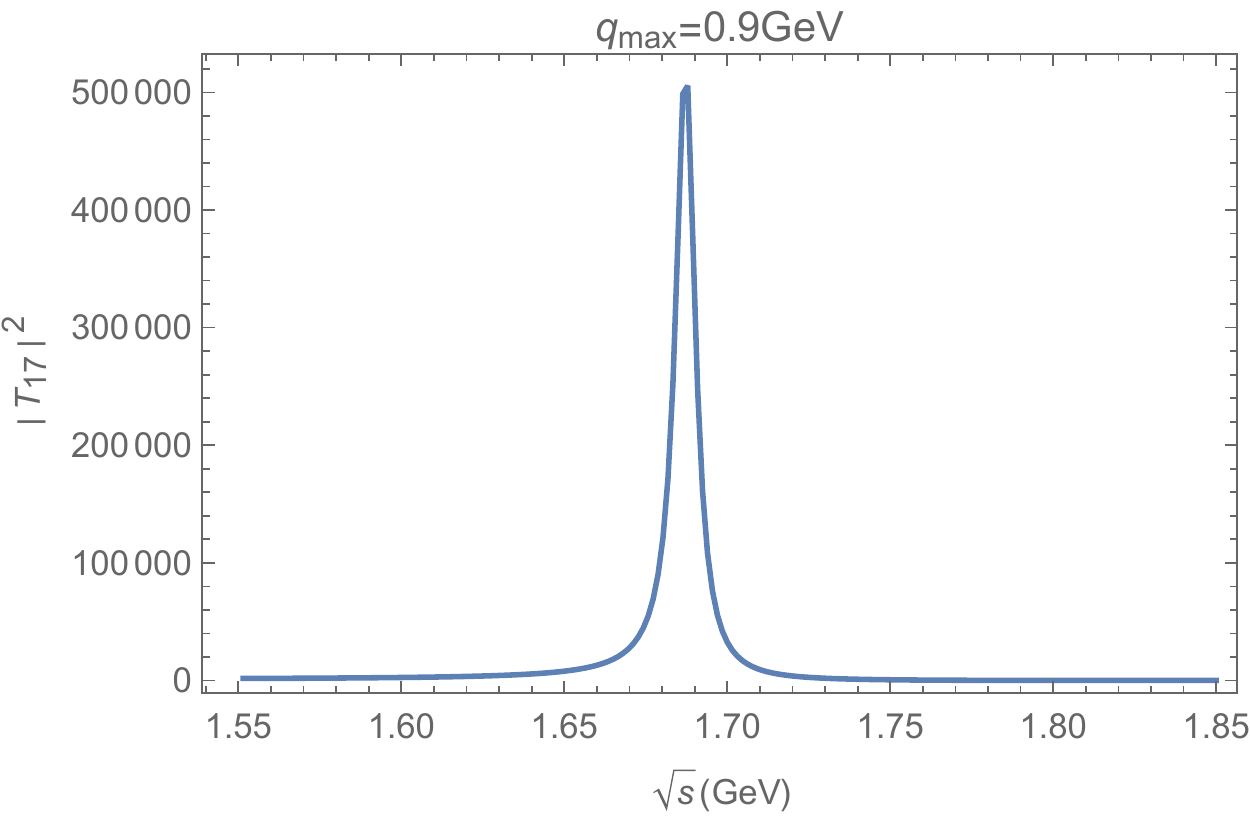}
  \includegraphics[scale=0.5]{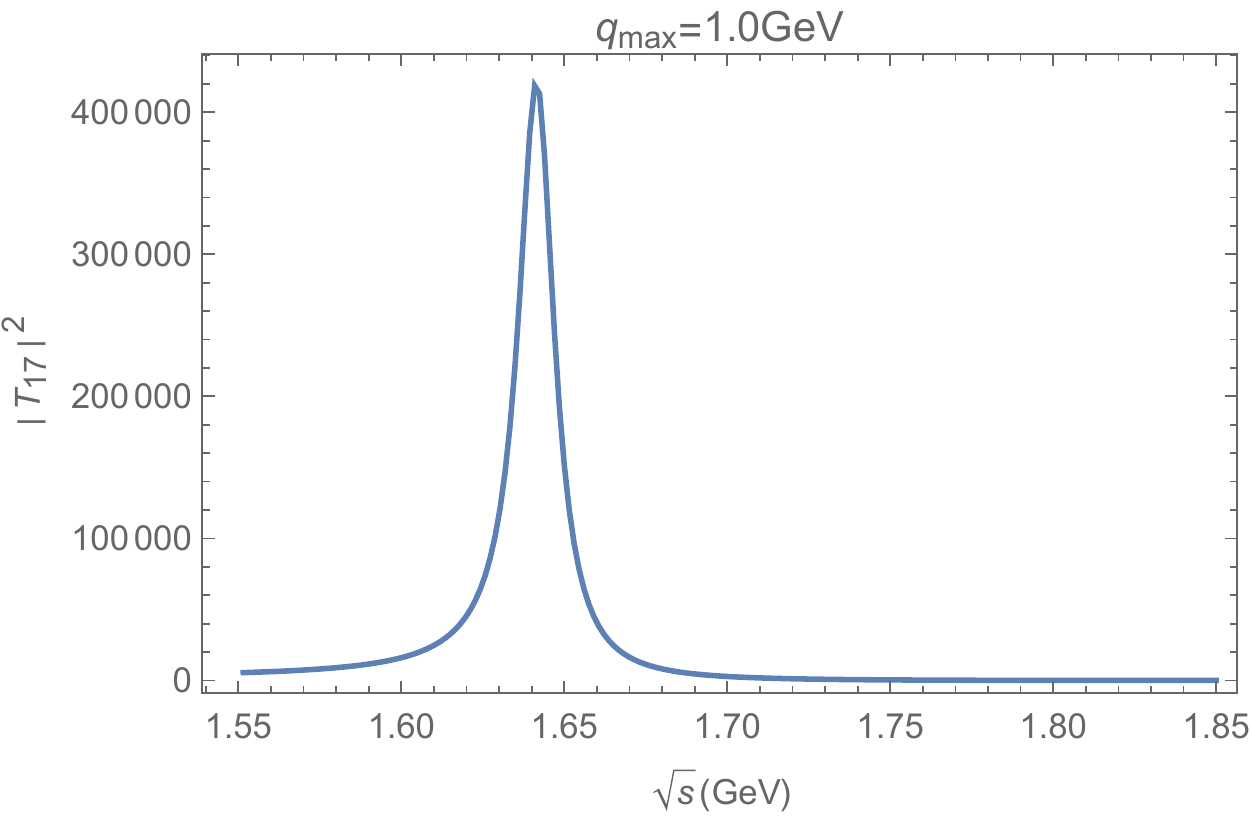}
  \caption{$|T_{17}|^2$ for $\Lambda=0.9 GeV$ and different $q_{max}=0.9,1.0 GeV$}\label{fig:t17}
\end{figure}

\begin{figure}
  \centering
  \includegraphics[scale=0.5]{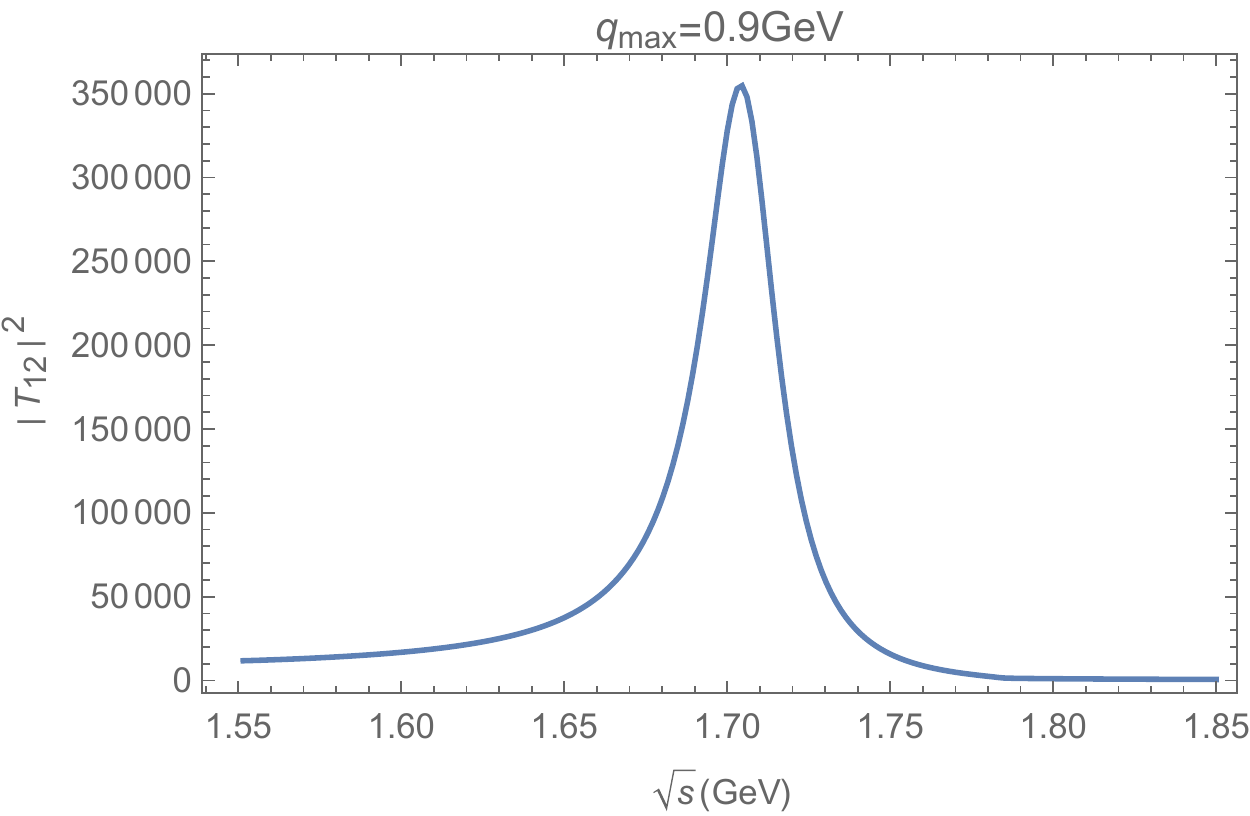}
  \includegraphics[scale=0.5]{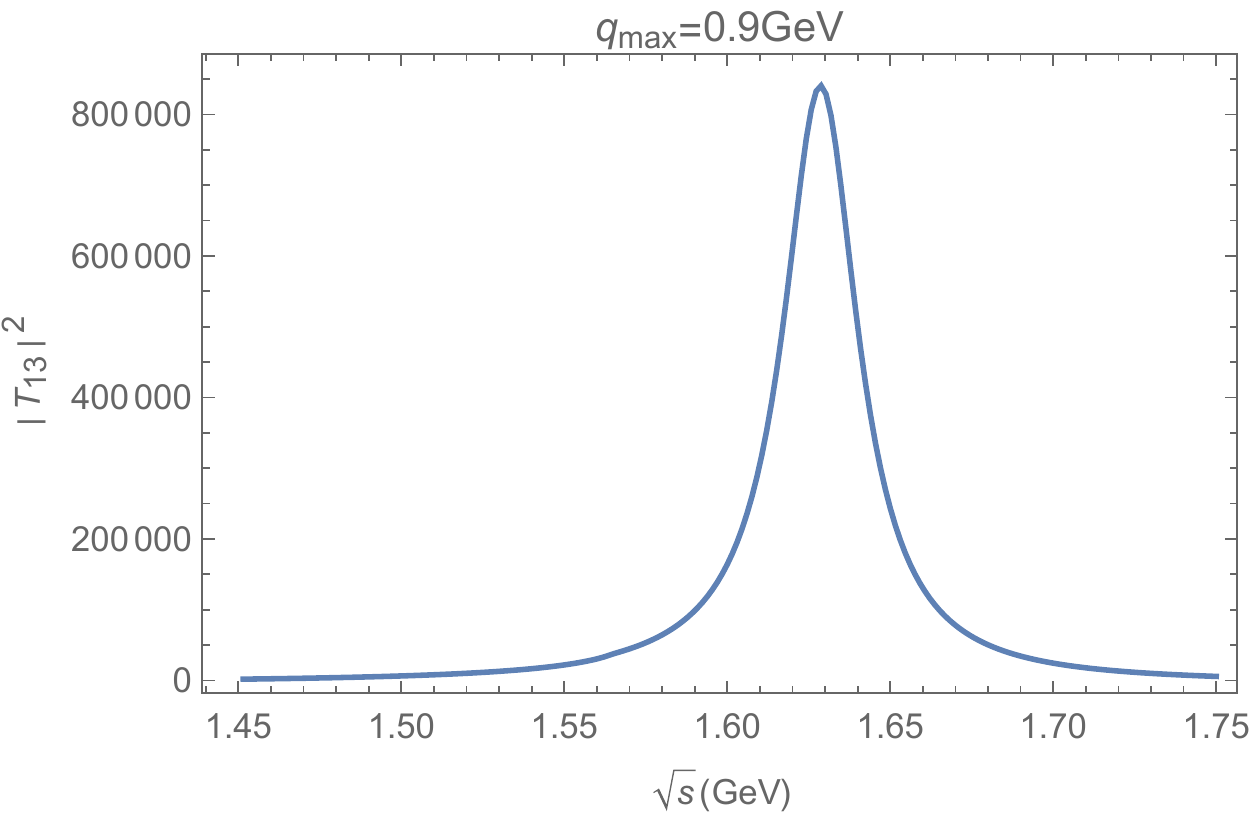}
  \caption{$|T_{12}|^2$ and $|T_{13}|^2$ for $\Lambda=0.9 GeV$ and $q_{max}=0.9 GeV$}\label{fig:t123}
\end{figure}

For the $K^*\bar{K}^*-K\bar{K}$ system, the $|T|^2$ is showed in Fig.~\ref{fig:t16}.
The resonance is $\sqrt{s}=1.76-0.015i GeV$ for $q_{max}=0.9 GeV$ and 
$\sqrt{s}=1.75-0.022i GeV$ for $q_{max}=1.0 GeV$. Compared to the single channel,
turning on the $K\bar{K}$ channel make the resonance move up along the real axis.
And the situation is similar for $K^*\bar{K}^*-\pi\pi$ system,
 see Fig.~\ref{fig:t17}.
The resonance is at $\sqrt{s}=1.69-0.004i GeV$ for $q_{max}=0.9 GeV$ and 
$\sqrt{s}=1.64-0.007i GeV$ and $q_{max}=1.0 GeV$. We find the width of the
resonance in $K^*\bar{K}^*-\eta\eta$ system is about equivalence to that
in $K^*\bar{K}^*-\pi\pi$ system, and about $2$\~{}$3$ times smaller than that
in $K^*\bar{K}^*-K\bar{K}$ system. For comparison, we also show the $|T|^2$
for $K^*\bar{K}^*-\rho\rho$ and $K^*\bar{K}^*-\omega\omega$ system,
see Fig.~\ref{fig:t123}.
The resonance is about $\sqrt{s}=1.70-0.014i GeV$ for $K^*\bar{K}^*-\rho\rho$
and  $\sqrt{s}=1.63-0.013i GeV$ for $K^*\bar{K}^*-\omega\omega$ with the cutoff
$q_{max}=0.9 GeV$. It is interesting that turning on the $\omega\omega$ channel
makes the pole to move down, recall that the pole is at $1.66 GeV$ for $K^*\bar{K}^*$
single channel. Finally, we turn on all channels
\begin{figure}
  \centering
  \includegraphics[scale=0.5]{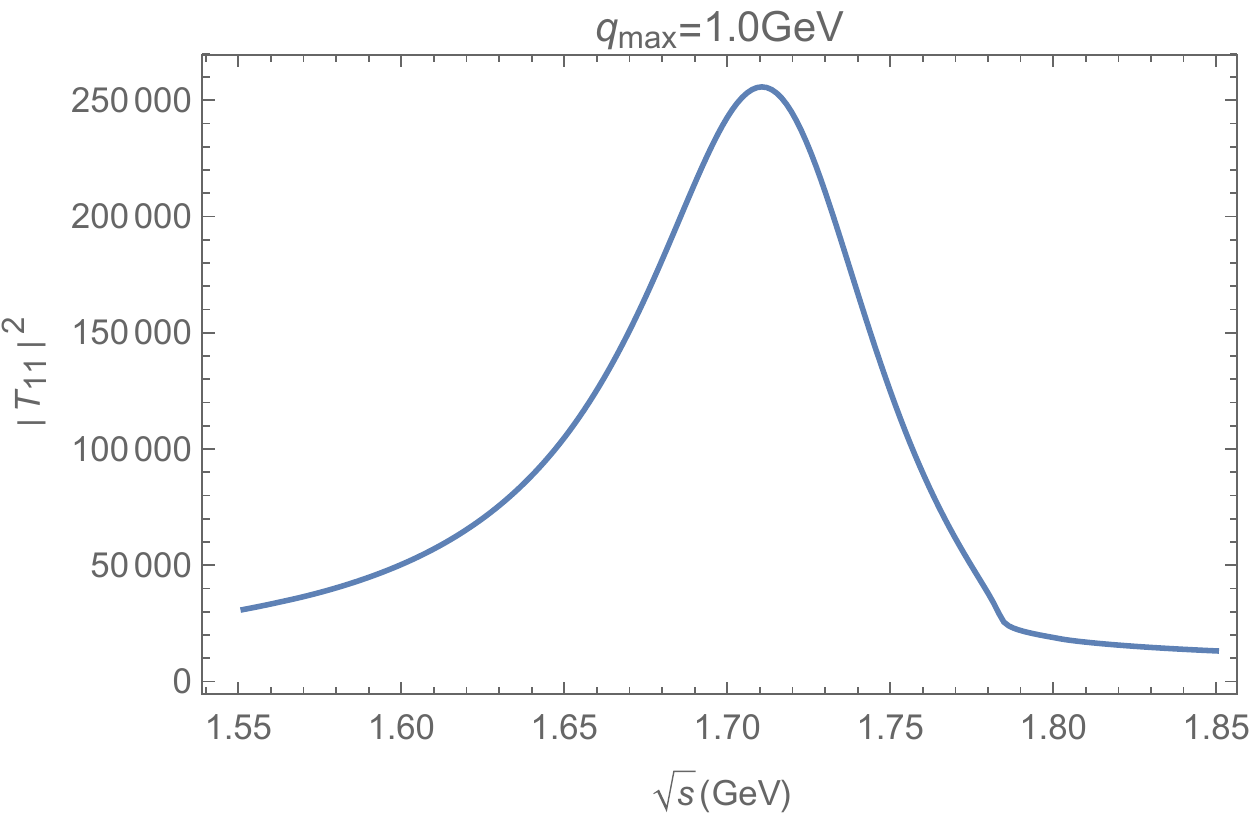}
  \includegraphics[scale=0.5]{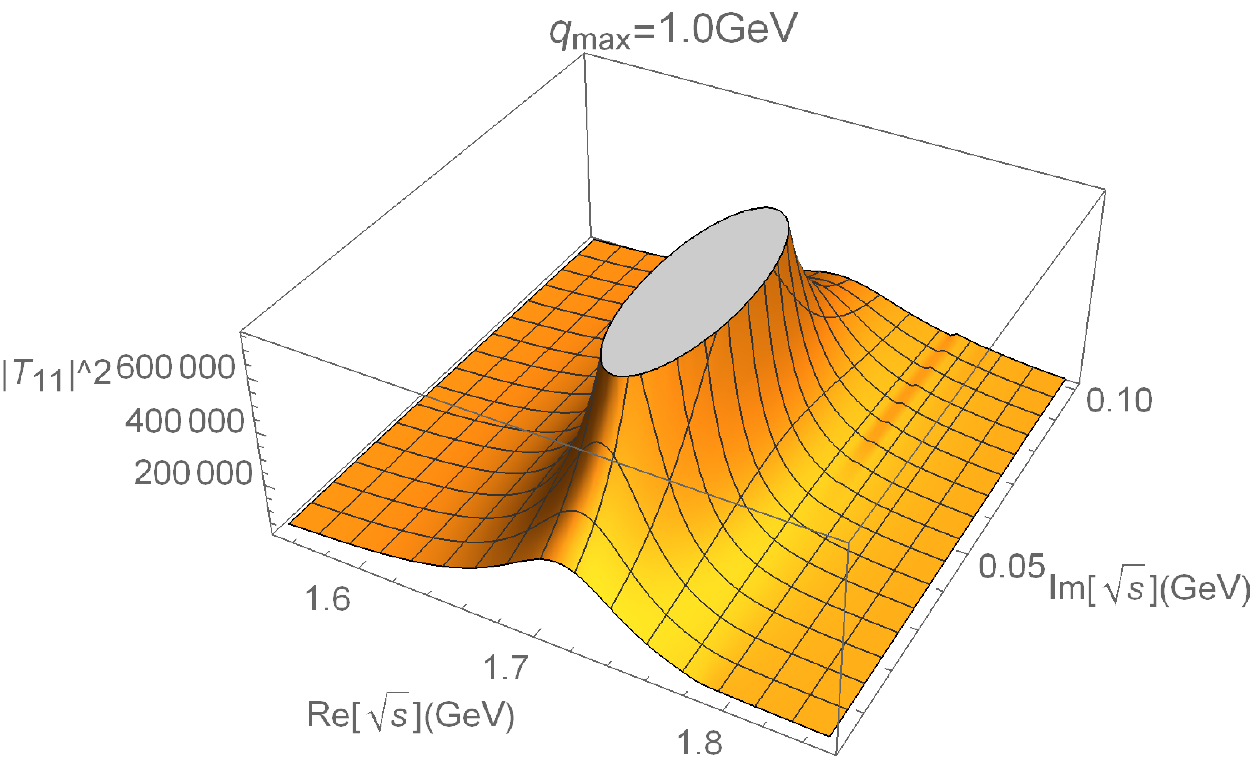}
  \caption{$|T_{11}|^2$ for $\Lambda=0.9 GeV$ and $q_{max}=1.0GeV$ in 2-dim and 3-dim}\label{fig:t11all}
\end{figure}
and show the $|T_{11}|^2$ in 2-dim and 3-dim with $q_{max}=1.0 GeV$ in Fig.~\ref{fig:t11all}. The position
of the resonance is listed in Table~\ref{tab:pole} for different cutoffs.
\begin{table}
  \centering
  \[\begin{array}{cccccc}
    \hline
    q_{max}(GeV) & 0.7 & 0.8 & 0.9 & 1.0 & 1.1 \\
    \hline
    Pole(GeV) & 1.77-0.015i & 1.75-0.028i & 1.73-0.035i & 1.72-0.045i & 1.70-0.053i \\
    \hline
  \end{array}\]
  \caption{The resonance pole for different cutoffs}\label{tab:pole}
\end{table}
We find that the real part of resonance is about $1710 MeV$ and the width is about
$100 MeV$, in fairly agreement with $f_0(1710)$, whose mass is $1704\pm 12MeV$
and width is $123\pm 18MeV$~\cite{PDG}. This means that the $f_0(1710)$ can be dynamically 
generated by mesons scattering. And we also find the lower resonance at
$1.46-0.012iGeV$ for $q_{max}=1.0GeV$ and $1.48-0.008iGeV$ for $q_{max}=0.875GeV$.
Compared to our previous work~\cite{Wang:2019niy}, the resonance move down a little
alone the real axis, which is $1.52-0.009iGeV$ for $q_{max}=1.0GeV$ and
$1.53-0.005iGeV$ for $q_{max}=0.875GeV$ in~\cite{Wang:2019niy}.

For the unitary coupled channel approach, we can calculate the ratio of decay width via~\cite{PDG}
\begin{equation}
  \Gamma_{R\to a} = \frac{|\tilde{g}_a|^2}{M_R}
  \rho_a(M^2_R).
\end{equation}
with $\tilde{g}_a = \mathcal{R}_{ba}/\sqrt{\mathcal{R}_{bb}}$ and
$\rho_a$ the two-body phase space. The residues may be calculated via
\begin{equation}
  \mathcal{R}_{ba}=-\frac{1}{2\pi i} \oint ds \mathcal{M}_{ba}.
\end{equation}
The branching ratios obtained this way are
\begin{align}
  \frac{\Gamma(f_0(1710)\to \pi\pi)}{\Gamma(f_0(1710\to K\bar{K}))}
  &=0.289\pm 0.092\; (0.23\pm 0.05), \\
  \frac{\Gamma(f_0(1710)\to \eta\eta)}{\Gamma(f_0(1710\to K\bar{K}))}
  &=0.294\pm 0.048\; (0.48\pm 0.15).
\end{align}
where the values of the PDG~\cite{PDG} are given in the brackets at the end of each equation for comparison. In fact the deviation
from various collaborations is much larger than the PDG range: the value of 
$\Gamma(f_0(1710)\to \pi\pi)/\Gamma(f_0(1710\to K\bar{K}))$ is 
$0.64 \pm 0.27 \pm 0.18$ in~\cite{Lees:2018qrk}, 
$0.41_{-0.17}^{+0.11}$ in~\cite{Ablikim:2006db},
$0.2 \pm 0.024 \pm 0.036$ in~\cite{Barberis:1999cq},
$0.39 \pm 0.14$ in~\cite{Armstrong:1991ch} and 
$0.32 \pm 0.14$ in~\cite{Albaladejo:2008qa}. The value of 
$\Gamma(f_0(1710)\to \eta\eta)/\Gamma(f_0(1710\to K\bar{K}))$ is 
$0.48 \pm 0.15$ in~\cite{Barberis:2000cd} and
$0.46_{-0.38}^{+0.70}$ in~\cite{Anisovich:2001ay}.
And for the radiative decays of $J/\psi$ in Table~\ref{tab:rdj}
from PDG~\cite{PDG},
\begin{table}[htb]
  \centering
  \[\begin{array}{|c|c|}
    \hline
    \text{Mode} & \text{Fraction}(\Gamma_i/\Gamma) \\
    \hline
    \gamma f_0(1710) \to \gamma K\bar{K} & 9.5^{+1.0}_{-0.5} \times 10^{-4} \\
    \hline
    \gamma f_0(1710) \to \gamma \pi\pi & 3.8^{+0.5}_{-0.5} \times 10^{-4} \\
    \hline
    \gamma f_0(1710) \to \gamma \eta\eta & 2.4^{+1.2}_{-0.7} \times 10^{-4} \\
    \hline
  \end{array}\]
  \caption{Some modes of radiative decays of $J/\psi$}\label{tab:rdj}
\end{table}
all we can say is that the partial decay widths of $f_0(1710) \to \pi\pi$ and
$f_0(1710) \to \eta\eta$ are similar to be around $1/3$ of 
$f_0(1710) \to K\bar{K}$, which are compatible with our results.


In summary, we extend the coupled channel interaction of nonet of vectors by including channels of the octet of
pseudo-scalars in addition using the unitary coupled-channel approach. The pole near the $K^*\bar{K}^*$
threshold remains to be there with mass and width consistent with PDG values of $f_0(1710)$. Meanwhile we deduce the partial decay widths of $f_0(1710) \to K^*\bar{K}^* \to \pi\pi,K\bar{K},\eta\eta$ in the approach as well as hadronic triangle loop approach for hadronic molecule. In both cases, the results agree with that of $f_0(1710)$ in PDG. We can conclude
that the properties of $f_0(1710)$ are consistent with the $K^*\bar{K}^*$ molecule state.

\section*{Acknowledgments}

We thank useful discussions and valuable comments from Feng-Kun Guo, Ulf-G. Mei{\ss}ner and Jia-Jun Wu. 
This work is supported by the NSFC and the Deutsche Forschungsgemeinschaft (DFG, German Research
Foundation) through the funds provided to the Sino-German Collaborative
Research Center TRR110 “Symmetries and the Emergence of Structure in QCD”
(NSFC Grant No. 12070131001, DFG Project-ID 196253076 - TRR 110), by the NSFC 
Grant No.11835015, No.12047503, and by the Chinese Academy of Sciences (CAS) under Grant No.XDB34030000.


\end{document}